\documentstyle[aps,prl,multicol]{revtex}
\input{epsf}

\draft
\begin{document}
\begin{multicols}{2}
\narrowtext

{\bf Comment on ``No enhancement of the localization length for two
  interacting particles in a random potential''}

 In a recent letter \cite{Roe97} R\"omer and Schreiber report
on numerical calculations that led them to conclude that the 
previously observed enhancement \cite{She94,Imr95,Fra95,Wei95,Opp96} of the 
localization length $L_2$ of two interacting particles (TIP) vanishes 
in the thermodynamic limit. Such a claim (i) is in conflict with the
scaling theory of localization, (ii) ignores
a consistent picture from a wealth of published
numerical data and analytical investigations, and (iii) directly
contradicts new numerical results obtained with a Green function
method \cite{Opp96} that is well adapted to the problem under study.

(i) R\"omer and Schreiber considered system sizes $L$ between $80$ and
$360$ for their extrapolation. For $L=360$ they find $L_2\approx 18$,
i.e. $L_2$ exceeds the one--particle localization length
$L_1\approx 12$ significantly. How can these extended states shrink
upon increasing the system size from $20L_2$ to, say, $25L_2$?
According to Thouless' scaling picture,
the energy shift associated with such an increase of $L$ amounts 
to a negligible fraction of the mean level spacing, and no significant 
effect can be expected.

(ii) Let $n_1$ and $n_2$ be the sites of the two particles.
Shepelyansky showed \cite{She94} that, {\it for a small fraction
($\approx L_1/(2L)$) of states}, the
center of mass $(n_1+n_2)/2$ is extended over $L_2$, while 
$|n_1-n_2| \lesssim L_1$. 
Alternative and complementary derivations are provided by Imry's 
variant\cite{Imr95} of the Thouless scaling picture, and by an 
effective $\sigma$ model \cite{Fra96}, both not restricted to 
finite $L$.  Three different numerical methods have 
confirmed this effect. First, extended states of the 
type described above were obtained by exact diagonalization \cite{Wei95} 
of the TIP Hamiltonian in a ring geometry. Second, the delocalization of the
center of mass was studied in Ref.~\cite{Opp96} by calculating the
TIP Green function $\langle n_1 n_2|G_2(E)|n_1' n_2'\rangle$
for $n_1=n_2=0$ and $n_1'=n_2'=L$, i.e. {\it pair transfer} from $0$ to $L$.
Third, the transport of {\it one} particle through a
relatively small sample containing the
second particle was investigated \cite{Fra95} with the transfer matrix
method. They all reach a common conclusion: For an on-site interaction 
$U=1$ and $L_1 \approx 11$, one has $L_2\approx 25 > L_1 $. Comparable 
results were obtained in {\it infinite} systems \cite{Fra95} with a 
prescribed maximum separation $|n_1-n_2|$ (``bag model'').

(iii) R\"omer and Schreiber used transfer matrix methods 
(similar to those in Ref.\cite{Fra95}), and concluded that 
$L_2 \to L_1$ when $L \to\infty$. In Fig.~1, 
we compare data from Ref.~\cite{Roe97} with results from
the Green function method ($U=1, L_1\approx 11$).
The latter consistently gives $L_2 
\approx 25 \approx 2 L_1$ up to the maximum size we considered, 
$L=1000$. Without explaining the data of Ref.~\cite{Roe97} in detail,
we emphasize the following important problem:
The $L^2$ entries in the transfer matrix correspond to the 
$L^2$ functions $\langle 0n_2|G_2(E)|Ln_2'\rangle$. Most of these
characterize single particle transport (decay scale $L_1$), only those
with $n_2\lesssim L_1$, $n_2'\gtrsim L-L_1$ describe pair transport
(decay scale $L_2/2$~\cite{Opp96}, with the conventions of
Ref.~\cite{Roe97}). To distinguish these two scales one has to either
suppress single particle transport (Green function method, bag model)
or ensure that $L_2\gg 2L_1$ (as in Ref.~\cite{Fra95}). Unfortunately,
Ref.~\cite{Roe97} meets neither of these requirements, and it is
unclear, which scale is actually measured.

In conclusion, the central claim in Ref.~\cite{Roe97} is due to a
serious misinterpretation, and the localization length $L_2$ of the 
interaction-assisted states is {\it independent} of $L$, as expected 
on physical grounds.

\begin{figure}

\epsfxsize=2.7in
\centerline{
\epsfbox{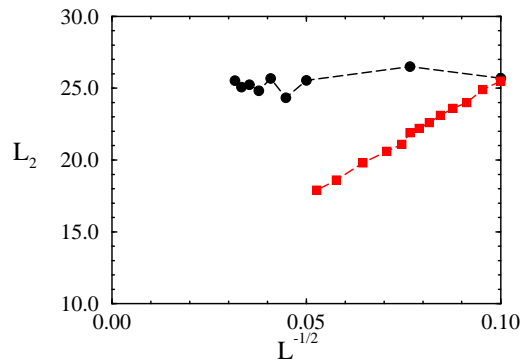}
}

\caption{Data for $L_2$ from Ref.~\protect\cite{Roe97} (squares) and
  the Green function method (circles)
  versus $L^{-1/2}$. We thank F. v. Oppen and T. Wettig for their
  computer program.}
\label{fig1}
\end{figure}

\noindent
Klaus Frahm$^1$, Axel M\"uller--Groeling$^2$, Jean--Louis Pichard$^3$,
and Dietmar Weinmann$^4$ \\
\\
$^1$Laboratoire de Physique Quantique, Universit\'e Paul Sabatier, 
F--31062 Toulouse, France\\
$^2$Max--Planck--Institut f\"ur Kernphysik, D--69029 Heidelberg, 
Germany \\
$^3$SPEC, CEA-Saclay, F--91191 Gif--sur--Yvette, France \\
$^4$Institut f\"ur Physik, Universit\"at Augsburg, D--86135 Augsburg,
Germany \\
\\
PACS numbers: 72.15.Rn, 71.55.Jv, 72.10.Bg

\end{multicols}
\end{document}